# Toy Models and Statistical Mechanics of Subgraphs and Motifs of Genetic and Protein Networks

S. Bumble, Department of Physics, Philadelphia Community College,
Philadelphia, PA 19130

**Abstract**

Theoretical physics is used for a toy model of molecular biology to assess conditions that lead to the edge of chaos (EOC) in a network of biomolecules. Results can enhance our ability to understand complex diseases and their treatment or cure.

Network motifs in biological networks are recurring, significant patterns of interconnections. They have excited more interest lately for their function in determining gene expression and for their basic computational design principles. Thus, earlier work[1,2] that had been done in other applications of statistical mechanics is recalled and applied to their possible use in biological networks. Such work did consider a gas adsorbed on a triangular lattice as an order-disorder problem of statistical mechanics. Lateral interactions were examined between nearest and next-nearest neighbors to find the region where the isotherm becomes critical. Subgraphs of a triangular network include the point, the bond, the triangle and the rhombus (Figure 1). Each of these was examined as representative of the whole network and conditions leading to critical isotherms or the EOC were considered. These were then characterized by their chemical potential (or eigenpotential) for specific values of the interactive energy and coordination number for each subgraph. The eigenpotential is proposed as defining the structure of the motif and as the communication or signaling entity with other motifs in other networks that lead to expressions of the biological networks.

A paper in the literature[1] studies the adsorption of a gas on to a solid lattice using statistical mechanics as formulated by Hijmans and DeBoer[3]. In this method the lattice is broken up into basic subfigures, such as the point, bond, triangle and rhombus. Constraining relations of normalization, consistency and equilibrium among these basic figures are maintained. By this method, isotherms of pressure versus coverage can be calculated and plotted for each degree of approximation. The present paper assumes that the reader is familiar with the above treatment and continues the method for the rhombus after the three constraining sets of relations were used and selected for the probability of occurrence of each degree of occupation of the rhombus.

Consider the subgraph below. It has six different energies of interaction. Four of them are the same when the proteins on sites 1 through 4 are identical. They are 1—2, 2—3, 3—4, and 4—1. Bonds 2—4 and 1—3 are internal bonds and they are different from each other as the distance 1—3 is greater than the distance 4—2. They are also different from the bonds 1—2, 2—3, 3—4, and 4—1. When this motif is placed into its global perspective the 1 and 3 proteins are next-nearest neighbors, whereas the 2 and 4 proteins are nearest neighbors. Now suppose the four sites are empty and will be filled by a pool of indistinguishable proteins that are in a reservoir. They will observe site preference when they occupy sites of the motif. Let us consider each step as if they proceed from the completely empty motif towards the completely occupied state.

Using statistical mechanics[2], we will find that proceeding from the state of being completely unoccupied to any state of occupancy can be formulated by the equation shown below:

(1) $\quad m = p_{is}/p_o$

where $p_o$ is the probability of the completely empty rhombus and $p_{is}$ is the probability for the rhombus where i sites are occupied in the s manner. Also,

(2) $\quad m = (x)^g(x')^h(y)^r(y')^s(z)^t$

where x is for occupied sites 1 and 3, x' is for occupied sites 2 and 4, y is for interaction 1—3, y' is for interaction 2—4 and z is for interactions 1—2, 2—3, 3—4 or 4—1. In equation 2, g is the number of sites in the equivalent positions 1 and 3 that are occupied, h is the number of sites in the equivalent position 2 and 4 that are occupied, r is for the bond formed in the position 1—3, s is for the bond formed in the position 2—4 and t is the number of bonds in the 1—2, 2—3, 3—4 and 4—1 positions. The values for the unknowns are found through complex algebra to be:

(3) $\quad x = K^2\tau,\ x' = K\tau,\ y = \sigma/\tau\ y' = C,\ z = C'$,

where $\tau$ and $\sigma$ are functions of K, C and C'. The superscripts in equation 1 can be found by simply counting in the diagrams but the values for K, $\tau$, and $\sigma$ must be found algebraically. Also the pathways between rhombi of different occupancy can then be found.

Symbolizing an occupied site by + and an unoccupied site by o, we then find the occupied sites as

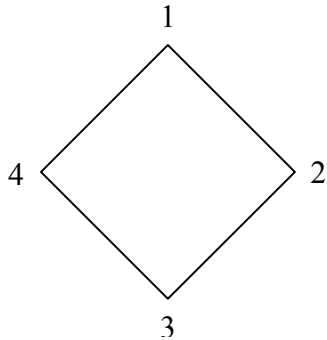

Figure 1

An occupied site is indicated by + and an unoccupied site by o. We then find the occupied sites as

$$\phantom{d_o = }\ 1234$$
$$d_o = oooo$$

(4) $\quad d_{11} = +ooo = K^2\tau$

(5) $\quad d_{12} = o+oo = K\tau$

(6) $\quad d_{22} = +o+o = (K\tau)^2 C' = C'K^2\tau 2$

(7) $\quad d_{212} = ++oo = (K^2\tau)(K\tau)(\sigma/\tau) = K^3\tau\sigma$

(8) $\quad d_{211} = o+o+ = (K^2\tau)^2 C = CK^4\tau^2$

(9) $\quad d_{31} = ++o+ = (K^2\tau)^2(K\tau)(\sigma/\tau)^2 C = CK^5\tau\sigma^2$

(10) $\quad d_{32} = +++o = (K^2\tau)(K\tau)^2(\sigma/\tau)^2 C' = C'K^4\tau\sigma^2$

(11) $\quad d_4 = ++++ = (K^2\tau)^2(K\tau)^2(\sigma/\tau)^4 CC' = C'CK^6\sigma^4$

It would seem from the above that symmetry is obeyed except for the symmetry between $d_{22}$ and $d_{211}$. This is false because one contains the short diagonal of the rhombus while the other contains the long diagonal of the rhombus.

There are six pathways between the probabilities of rhombi with succession occupation numbers as follows:

$$p_o \rightarrow p_{12} \rightarrow p_{22} \rightarrow p_{32} \rightarrow p_4$$
$$p_o \rightarrow p_{12} \rightarrow p_{212} \rightarrow p_{32} \rightarrow p_4$$
$$p_o \rightarrow p_{12} \rightarrow p_{212} \rightarrow p_{31} \rightarrow p_4$$
$$p_o \rightarrow p_{11} \rightarrow p_{212} \rightarrow p_{32} \rightarrow p_4$$
$$p_o \rightarrow p_{11} \rightarrow p_{211} \rightarrow p_{31} \rightarrow p_4$$
$$p_o \rightarrow p_{11} \rightarrow p_{212} \rightarrow p_{31} \rightarrow p_4$$

and the factors that provide the transitions between these probabilities are readily seen from the set of equations that precede this last set of equations. The word explosion in reference 16 can be expressed better by the word nucleation. The factors for the transitions in the last set of equations will be seen to be $\lambda = e^{u/kT}$ when each new protein appears on the rhombus and $C = e^{-w/kT}$, appears when nearest-neighbor proteins interact with each other, while $C' = e^{-w'/kT}$, appears when next-nearest neighbors react with each other. Here u is the chemical potential, T is the absolute temperature, k is the Boltzmann constant and w and w' are the interaction energies of the respective neighbors.

In order to obtain the foregoing, the assumptions had to be made that $C'-1 = 0$, although in certain cases, $C'-1$ was as high as 0.04. Also, the assumption was made that $K-1 < 1$, which indeed was always obeyed when interactions were attractive rather than repulsive. Those who wish to see results when these assumptions were not made can refer to the literature[1].

Now the normalizing and equilibrium relationships for each subfigure and the consistency relationships among all the subfigures lead to similar values for the variables s and t.

The point, bond and triangle are subfigures of the rhombus and are related through consistency relations. Thus, the probability of occupation of these subfigures are a function of K, $\sigma$ and $\tau$. Also, the normalizing relations of all occupation states of the subfigures are also in effect and so we can obtain values for $b_1$ and $b_2$, which are the probabilities of the sites 1, and both sites 1 and 2, resp., to be occupied divided by the probability of both sites 1 and 2 to be empty.

1 2
o-o

Now the fraction of sites occupied n the whole network is $\Theta$ so that
$$\Theta = (b_1 + b_2)/(1 + 2b_1 + b_2)$$
Leads to the polynomial:
$$K^4 - a_1 K^3 + a_2 K^2 - a_3 K + a_4 = 0,$$
where,
$$a_1 = [(C'+1)-(3C'+2)\Theta]C + C'(2 -3\Theta)]/C(1-2\Theta)$$
$$a_2 = [1 + (((2C'+3)-2\Theta(3C'+2))/C(1-2\Theta))/]C'$$

$$a_3 = [((2C'+1)-(4C'+1)\Theta]C+(C')^2(1-3\Theta))/C^2(1-2\Theta)]C'$$
$$a_4 = (C')^3/C^2$$

and the isotherm equation can be derived from the quartic equation, as

$$K = ((\exp(u/kT))(1-\Theta)/\Theta))^{\wedge}1/Z.$$

And then the plots of $e^{u/kT}$ versus $\Theta$ can be made for varying values of C and Z. We then adjust these two parameters until the isotherm becomes critical. We then find that when the subfigure (or subgraph) becomes larger, the isotherm becomes "stiffer". Thus the EOC is obtained at higher C and/or Z, but then becomes less prone to cross over to chaos. Thus functionality is more robust. Figure 2 is a graph of the isotherms when c = 3.0 and Z = 6. Notice the large loops for the bond isotherm, the slight loops for the triangle isotherm and the fact that the rhombus isotherm is just about at its critical point. The exact value for the critical isotherm when Z = 6 is c = 3.001.

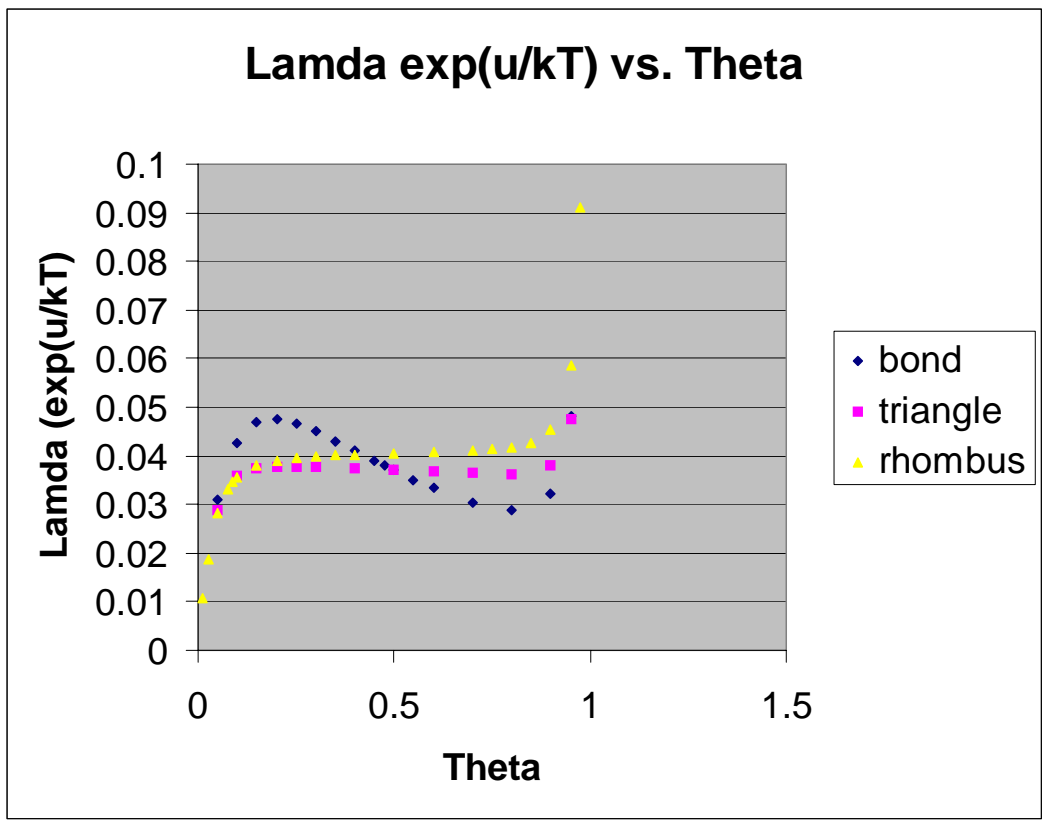

**Figure 2**

**Discussion**

In a working paper by Stuart A. Kauffman et alia[4] of the Santa Fe Institute[6], the interactions within a small group of people in a business environment is explored for the connections between (externalities) that are tuned independently. By controlling which groups have externalities with which other groups, the topology of the problem can be manipulated-the web of interactions within and between groups. Optimization of this system with time leads to a border of chaotic behavior. This model has been adopted in the present study using some statistical mechanics and thermodynamic relations from physical chemistry and applied in the field of molecular biology.

When two phases are in equilibrium, their chemical potentials are equal. If a motif or a module is in its critical region, its chemical potential is constant and it is at its EOC. We define $\lambda_c$ which is $\exp(u_c/kT)$ for a particular temperature and call it an eigenpotential. We can now find the eigen potential for a motif or module signified by the geometrical figure of the point, bond, triangle or rhombus and have delineated the method for obtaining it for the hexagon, etc. However each of these eigenpotentials are for $(\lambda_c)_{T, w, Z} = (\exp(u_c/kT))_{T,w,Z}$, at a particular temperature, interaction energy and coordination number. If we consider this motif or module as a phase with a particular critical eigenpotential, it would be in equilibrium with another phase that has the same eigenpotential. We know that if these motifs or eigenpotentials are within a network of genes or proteins confined within a chromosome then the motif or module may be very close to another motif or module that can have the same eigenpotential. It may also be true that this critical motif or module of constant chemical potential is able to send a signal to another biochemical cluster and modulate its composition and shape to correspond to the signal sender and form either a receptor, or similar geometric molecular form, to the signal sender so that the motif or module is propagated in a new network and is an emissary for a functionality of the global set of modules or motifs.

Some of these questions concern the Gibbs Phase Rule. Do we consider each distinguishable protein (or component) a different phase? Also, is each vertex or position of the protein distinguishable as well as each bond or interaction between adjacent proteins? These questions arise when we attempt to place this motif into a global concept. The other questions of importance are what factors make the system difficult or easy to reach the system's EOC, which facilitates functionality.

A paper by Sear[5] indicates that motifs containing many proteins may not exist because of all the coding necessary for its existence. However, the present paper presupposes the existence of at least a quartet of associated proteins. Another paper by Sole[6], states that real proteome maps when tuned close to a sharp transition point, separating a highly connected graph from a disconnected system, display scale-free hierarchical organization, behave as small worlds and exhibit modularity. These results can occur for cellular networks as a result of a duplication-diversication process.

Motifs are a set of genes or gene products with specific molecular functions arranged together so that they perform some useful behavior but they are not separable from the rest of the system[7]. Motifs are only part of a systems-level function (such as a feedback loop or a logical operation). Motifs that were widely used for sequence analysis were generalized to the level of networks.[8] Network motifs are likely to be also found on the level of protein signal networks. Finding a network motif in a new network may help explain what systems-level function the network performs, and how it performs it. One can unite related groups of motifs of different sizes into families termed motif generalizations[9]. This allows generalizations from small motifs to the larger complexes in which they occur, using efficient algorithms. Motifs can define universal classes of networks. They may be interpreted as structures that arise because of the special constraints under which the network evolved[10].

Modules are defined as discrete units of function separable from the whole. A module of genes may comprise common motifs arranged in new ways to produce different phenotypes. A module is a set of nodes that have strong interactions and a common function. They can be a set of design principles common both to circuits of human design and result from evolution[7].

The emergence of the hierarchical topology[11] through copying and reusing existing modules and motifs is a process reminiscent of gene duplication. Work has been described that visualize connections between subgraphs and for highlighting many similar subgraphs[12]. Evolution is a tinkerer not an engineer. There is a similarity between the creations of tinkerer and engineer[13]. We must understand the laws of nature that unite evolved and designed systems. Biological entities perform computations[14]. There is an evolutionary payoff placed on being able to predict the future. Complex organisms can better cope with environmental uncertainty because they can compute.

Graph theoretical quantities in biological networks consider degree distribution, degree-degree correlation function, assortativity, average clustering coefficient and the local clustering coefficient. Robustness under evolution, gene duplication, divergence and mutation are very important. Subgraphs in random networks and their influence on scaling[15] is discussed. Also,. the concept of protein families[16] in protein networks is emphasized.

**Conclusions**

The coordination number of each protein and gene in a network is very important to motif formation as are the interaction energies of the nearest and next-nearest neighbors in the network[17,18]. Also important is the chemical potential of the molecules of the motif. To understand the expression of these biological network's functions, as well as design efficient drugs, these factors can be very helpful. When one considers an assembly of molecules that defines a motif, a chemical potential can be assigned. This is the particular chemical potential at which the EOC is established. The EOC cluster is considered as a phase and is in equilibrium with other clusters of molecular assemblies whose chemical potential is equal to the original phase. This second motif phase grows in steps by preferential attachment of molecules, each of which is at the EOC for its size, until the size of the original or earliest motif is duplicated and this duplication can occur repeatedly in different networks until the final expression or phenotype is achieved. A vertex can be occupied by proteins that are rigidly docked in its preferred binding mode, occupied by proteins that first attach to the target and then fold[19],or occupied by unfolded proteins. In the last case, disease may follow, and the EOC will be more difficult to attain. Such proteins must not be allowed in a network.

Acknowledgement: I thank Laszlo Papp for applying his computer acumen.